\documentclass[tradiabstract]{aa}

\AddToHook{begindocument/before}{\usepackage{hyperref}}

\usepackage{graphicx}
\usepackage{amsbsy}
\usepackage{amsmath}
\usepackage{natbib}
\usepackage{color}
\usepackage{txfonts}
\usepackage{url}
\usepackage{bm}

\DeclareMathSymbol{\varOmega}{\mathord}{letters}{"0A}
\DeclareMathSymbol{\varSigma}{\mathord}{letters}{"06}
\DeclareMathSymbol{\varPsi}{\mathord}{letters}{"09}
\DeclareMathSymbol{\varPhi}{\mathord}{letters}{"08}
\DeclareMathSymbol{\varGamma}{\mathord}{letters}{"00}

\begin{document}

\title{Exploring the conditions for forming planetesimals by the streaming instability and planetary systems by pebble accretion}
\titlerunning{Conditions for streaming instability and pebble accretion}

\author{Anders Johansen\inst{1,2} \& Wladimir Lyra\inst{3}}
\authorrunning{Johansen \& Lyra}

\institute{$^1$ Center for Star and Planet Formation, Globe Institute,
University of Copenhagen, \O ster Voldgade 5-7, 1350 Copenhagen, Denmark, \\e-mail:
\url{Anders.Johansen@sund.ku.dk} \\ $^2$ Division of Astrophysics, Department of Physics, Lund University, Box 118, 22100 Lund, Sweden\\
$^3$ New Mexico State University, Department of Astronomy, PO Box 30001 MSC 4500, Las Cruces, NM 88001, USA }

\date{}

\abstract{The streaming instability and pebble accretion are two physical mechanisms with demonstrated potentials to drive, respectively, the formation of planetesimals and the growth of planetary systems containing a diverse range of planetary types. Here we explore the protoplanetary disc conditions in terms of turbulence strength, Stokes number and initial disc size that are needed to (i) form planetesimals by the streaming instability, (ii) form gas giant planets in cold orbits, (iii) form super-Earths and sub-Neptunes close to the star and (iv) form rocky planet embryos in temperate orbits. We identify an optimum  Stokes number range between ${\rm St}=0.01$ and ${\rm St}=0.03$ where all three planetary classes form and where the streaming instability is triggered for a slightly elevated pebble metallicity. Cold gas giants require a turbulence strength of at most $\delta=10^{-4}$ and furthermore need large initial disc sizes to benefit from a prolonged pebble flux; super-Earths and rocky planet embryos tolerate higher turbulence strengths similar to those measured for the vertical shear instability. A higher Stokes number of ${\rm St}=0.1$ is detrimental to the formation of cold gas giants due to the short-lived pebble flux. For Stokes numbers below ${\rm St}=0.003$, extremely low values of turbulence ($\delta<10^{-5}$) are required to form cold gas giants. We highlight how loss of gas to disc winds, reduction in the migration speed by thermal or dynamical torques or the presence of pressure bumps in the outer disc could increase the parameter space for the formation of cold gas giants. We derive analytically that the mass of the largest planetesimals formed by the streaming instability is of similar magnitude to the threshold mass beyond which pebble accretion becomes efficient, if planetesimals form in the earliest phases of protoplanetary disc evolution.}


\maketitle

\section{Introduction}
\label{s:intro}

The dense environment of a protoplanetary disc  facilitates collisional sticking of micrometer-sized particles consisting of refractory silicate/metal and volatile ices. Collisions driven by turbulence become ever-faster as the dust aggregates grow to larger sizes, until the collision speed reaches the threshold above which the internal rearrangement of monomers can no longer absorb the collision energy \citep{DullemondDominik2005,Brauer+etal2008}, so that the collision leads to fragmentation rather than sticking. The fragmentation threshold speed for silicate dust aggregates is generally found to lie at approximately 1 m/s \citep{Guttler+etal2010}, but its value may increase to 5--10 m/s at the high temperatures in the innermost regions of the protoplanetary disc due to the higher surface energy of dry silicates at elevated temperatures \citep{Kimura+etal2015}.

The sticking threshold of water ice has been measured to be very high, on the order of 10 m/s \citep{Gundlach+etal2015}. Experiments under realistic temperature conditions, applicable in the icy regions of the protoplanetary disc (temperature $T<170\,{\rm K}$), nevertheless show a dramatic drop in surface energy \citep{Gundlach+etal2018} and imply a fragmentation threshold speed for water ice that is similar to that of silicates \citep{MusiolikWurm2019}. Even more volatile CO$_2$ ice has also been experimentally probed to have a sticking threshold similar to that of silicates \citep{Musiolik+etal2016,FritscherTeiser2021}.

Reaching the fragmentation barrier requires that the aggregates remain porous
during their growth, since compact particles experience bouncing already at much
lower collision speeds
\citep{Zsom+etal2010,Arakawa+etal2023,DominikDullemond2024,QianWu2025}. Also, if monomers
are smaller than the usually assumed 1 $\mu$m, this would lead to higher
fragmentation thresholds \citep{Okuzumi+etal2012}. The fragmentation barrier
\citep{Birnstiel+etal2011}, despite its simplifications in terms of ignoring porosity evolution, has nevertheless emerged as a useful reference point for understanding growth from dust to pebbles in protoplanetary discs \citep{Jiang+etal2024,Tong+etal2025}. The Stokes number St of pebbles at the fragmentation barrier is found by equating the threshold speed for fragmentation, $v_{\rm f}$, with the turbulent collision speed, $v_{\rm t}=\sqrt{3 {\rm St}\, \delta} c_{\rm s}$, to obtain
\begin{equation}
  {\rm St} = \frac{1}{3} \frac{1}{\delta} \left( \frac{v_{\rm f}}{c_{\rm s}} \right)^2 \approx 0.01 \left( \frac{\delta}{10^{-4}} \right)^{-1} \left( \frac{T}{100\,{\rm K}} \right)^{-1} \left( \frac{v_{\rm f}}{1\,{\rm m/s}} \right)^2 \, .
  \label{eq:fraglim}
\end{equation}
Here, $\delta$ is a dimensionless measure of the strength of the turbulence
(related to the turbulent diffusion coefficient $D$ through  the scaling
$D=\delta c_{\rm s} H$ where $c_{\rm s}$ is the gas sound speed and $H$ is the
gas scale-height). A fragmentation threshold of $v_{\rm f}$$\sim$$1$ m/s and a
turbulence strength $\delta$$\sim$$10^{-4}$ are found to be in good agreement
with the mm-cm wavelength emission from pebbles in a range of observed
protoplanetary discs \citep{Jiang+etal2024,Tong+etal2025} and indicates a
nominal Stokes number of order ${\rm St}$$\sim$$0.01$. Extraction of the pebble Stokes
number from protoplanetary disc observations nevertheless comes with a large
uncertainty, due to among other things the assumed opacity of the
pebbles and their poorly constrained porosity \citep{Ueda+etal2025}.

The realization that particle growth in protoplanetary discs likely stalls at pebble
sizes -- historically achieved from a combination of dust collision experiments
\citep{BlumWurm2000}, computer simulations of dust coagulation including
fragmentation and radial drift \citep{DullemondDominik2005} and observations of
emission from pebbles in protoplanetary discs at mm-cm wavelengths
\citep{Testi+etal2003,Wilner+etal2005} -- inspired the development of the
streaming instability theory for planetesimal formation
\citep{YoudinGoodman2005,YoudinJohansen2007,JohansenYoudin2007}. Through the
action of the streaming instability, radially drifting pebbles will
spontaneously gather into dense filaments due to their drag force interaction with the background gas, forming 100-km-scale planetesimals from the gravitational contraction of the filaments \citep{Johansen+etal2007,Johansen+etal2015,Simon+etal2016}. However, the threshold metallicity, defined as the local pebble surface density relative to the gas surface density, needed for triggering the emergence of dense filaments depends both on the pebble Stokes number and the strength of the background turbulence \citep{Carrera+etal2015,Yang+etal2017,LiYoudin2021,Lim+etal2024}.

The presence of planetesimals, left-overs from the formation epoch of our Solar
System, in the asteroid belt between Mars and Jupiter inspired the
planetesimal-driven model of planet formation \citep{Safronov1969}. A
combination of planetesimal accretion and giant impacts between protoplanets
can broadly explain the formation of planets akin to Earth and Venus in mass over a time-scale of $\sim$$100$ million years \citep{ChambersWetherill1998,KokuboIda2000,Raymond+etal2004}, and the role of
giant impacts in the growth of terrestrial planets is supported by the need for
the Earth to suffer at least one impact in order to explain the mantle-like bulk
composition of the moon \citep{KokuboIda2007}.

In contrast, the accumulation time-scale of 100-km-scale planetesimals in the outer Solar System, where gas giants and  ice giants form, takes much longer than the life-time of the protoplanetary disc \citep{Thommes+etal2003,Levison+etal2010,JohansenBitsch2019,LorekJohansen2022,Kaufmann+etal2025}. Considering much smaller planetesimals (of $\sim$100 m scale rather than $\sim$100 km) yields higher accretion rates due to enhanced capture rate by gas \citep{Fortier+etal2013,Chen+etal2025}, but using such small planetesimals to drive planet formation is in conflict with the observed characteristic 100-km-scale of asteroids \citep{Bottke+etal2005,Johansen+etal2015}.

The pebble accretion theory was developed as a solution to the inefficiency of
planetesimal accretion (using nominal sizes of 100 km) in forming the cores of
giant planets within the life-time of the protoplanetary disc while avoiding 
the cores from migrating to the terrestrial planet zone
\citep{JohansenLacerda2010,OrmelKlahr2010,LambrechtsJohansen2012}. In pebble
accretion, gas drag greatly enhances the accretion cross section of mm-cm-sized
pebbles relative to larger planetesimals. High pebble accretion rates onto
planetesimals and small protoplanets
nevertheless require low turbulence strengths in the outer regions of the
protoplanetary disc where the cores of gas giants grow
\citep{LambrechtsJohansen2014,Ormel2017,Chambers2018,JohansenLambrechts2017,Gurrutxaga+etal2024,ZhaoMatsumura2025},
since these small bodies initially accrete in the 3-D regime where the local pebble density is set by the balance between sedimentation and turbulent diffusion. Such weak disc turbulence has indeed been inferred from observations of a high degree of pebble sedimentation in some protoplanetary discs
\citep{Pinte+etal2016,Villenave+etal2022}. However, inferring the level of
turbulence in protoplanetary discs is not straightforward and relies on many
assumptions about e.g.~the pebble sizes and gas surface densities. Also, the
turbulence level in protoplanetary discs likely varies from disc to disc,
between the inner and outer regions and over the evolutionary stage
\citep{Villenave+etal2023,Bosman+etal2023,Villenave+etal2025,Jiang+etal2025}.
Pebble accretion can also contribute significantly to the growth of rocky
planets, super-Earths and sub-Neptunes in the inner disc
\citep{Lambrechts+etal2019,Johansen+etal2021,Izidoro+etal2021,Nielsen+etal2025,Danti+etal2025},
but the growth rate by pebble accretion relative to the rate of planetesimal accretion and giant
impacts in the inner disc  depends on pebble sizes and turbulence levels 
\citep{YapBatygin2024}, as well as on the assumed population of planetesimals
that forms in the inner disc.

In summary, the operation of both the streaming instability and pebble accretion
have since their discovery been probed in great detail using computer
simulations. Particularly, the conditions for filament formation by the
streaming instability and efficient pebble accretion in terms of Stokes number,
turbulence strength and metallicity are now relatively well understood. These
conditions are nevertheless rarely explored in the combined light of driving both streaming instability and pebble accretion. The goal of this paper is therefore to provide a better map of the outcome of planet formation as a function of the range of plausible conditions in the protoplanetary disc. We believe that this endeavor gives important insights into the influence of the protoplanetary disc environment on the formation of planetary systems. 

The paper is organized as follows. In Section \ref{s:disc} we describe the protoplanetary
disc model that we use to calculate the conditions for the streaming instability
and the growth of planetary systems by pebble accretion. In Section \ref{s:streaming} we go on to
explore the conditions for the forming planetesimals by the streaming
instability at the water ice line. In Section \ref{s:planetesimals} we calculate the growth
time-scale of planetesimals formed by the streaming instability, highlighting a
new result that the characteristic mass of planetesimals formed by the streaming
instability is of similar magnitude to the threshold mass for efficient pebble
accretion. In Section \ref{s:pebble} we present synthetic populations of planetary systems formed by pebble accretion for a range of protoplanetary disc parameters. In Section \ref{s:map} we map the conditions for the streaming instability and the outcome of pebble accretion simulations for a large range of pebble Stokes numbers and turbulence levels. Finally, in Section \ref{s:summary} we briefly summarize the key findings of the paper.

\section{The protoplanetary disc model}
\label{s:disc}

We calculate the conditions in an evolving  protoplanetary disc at the orbital distance
of a planetesimal or a protoplanet using the analytical expressions derived in \cite{Gurrutxaga+etal2024}. That model uses a viscous $\alpha$-disc approach to the gas combined with an exact analytical expression for the evolution of the pebble surface density in response to gas drag and radial drift. We assume here that the product of Stokes number St and logarithmic pressure gradient $\chi=-\partial \ln P /\partial \ln r$ is constant. This yields a relatively constant Stokes number in the inner disc followed by a radially declining Stokes number in the exponentially tapered outer regions of the disc where the pressure gradient increases; such a decline was demonstrated in \cite{Gurrutxaga+etal2024} to give a good match to a more advanced coagulation model presented in \cite{Appelgren+etal2023} that took into account that pebbles in the outer disc are limited in growth by the drift barrier rather than by the fragmentation barrier. We note that we must ignore the explicit temperature dependence of fragmentation-limited pebble growth in equation (\ref{eq:fraglim}) in order exploit the analytical pebble flux expressions from \cite{Gurrutxaga+etal2024}.

\subsection{Evolution of gas and pebbles}

The ratio of pebble surface density to gas surface density $Z$ is, surprisingly, independent of distance to the star when the product ${\rm St}\,\chi \equiv {\rm St}_\chi$ is assumed constant, and takes the temporal form
\begin{equation}
  Z(t) = Z_0 T_{\rm e}^{-\frac{b_0}{2 (2-\gamma)}} \, ,
\end{equation}
where $Z_0$ is the initial surface density ratio, $\gamma=15/14$ is the
power-law exponent of the viscosity and $b_0=(2/3) {\rm St}_\chi/\alpha$ is a
constant that compares the radial drift speed of the pebbles (through ${\rm St}_\chi$) to the radial accretion speed of the gas (through $\alpha$). The dimensionless evolutionary time $T_{\rm e}$ is defined as
\begin{equation}
  T_{\rm e}=t/t_{\rm s}+1
\end{equation}
with $t$ representing physical time and $t_{\rm s}$ denoting the viscous time-scale over the initial disc size $R_1$,
\begin{equation}
  t_{\rm s} = \frac{1}{3 (2-\gamma)^2} \frac{R_1^2}{\nu_1} = 0.75\,{\rm Myr}\, \left( \frac{R_1}{100\,{\rm AU}} \right)^{13/14} \left( \frac{\alpha}{0.01}\right)^{-1} \left( \frac{L}{L_\odot}\right)^{-2/7} \, .
\end{equation}
Here, $\nu_1 = \alpha c_{{\rm s},1} H_1$ is the viscosity evaluated at the initial radius $R_1$. The luminosity scaling comes from adopting the irradiated temperature profile from \cite{Ida+etal2016}. The characteristic size of the disc $R_{\rm c}$ increases with time as
\begin{equation}
  R_{\rm c} (T_{\rm e}) = R_1 T_{\rm e}^{1/(2-\gamma)} \, .
  \label{eq:Rc}
\end{equation}

We have chosen to adopt an $\alpha$-disc model for the gas due to its relative
simplicity with only two free parameters governing its evolution: the
dimensionless viscosity $\alpha$ and the initial disc size $R_1$. The
$\alpha$-disc model captures well the accretion of gas onto the star, with a
speed proportional to the choice of $\alpha$
\citep{Hartmann+etal1998,Johansen+etal2019}. Angular momentum transport is
nevertheless in reality most likely not facilitated by turbulent viscosity, as
envisioned in the $\alpha$-disc approach, but rather by magnetocentrifugal disc
winds \citep{Bethune+etal2017,Mori+etal2019}. An analytical protoplanetary disc
evolution model including loss of angular momentum and mass by disc winds was
derived in \cite{Tabone+etal2022} and used to calculate pebble accretion growth tracks in
\cite{ZhaoMatsumura2025}. Other versions of disc wind models have also been derived \citep{Chambers2019}. Although a disc wind model is more realistic than the
$\alpha$-disc model, it comes at the expense of introducing two additional free
parameters ($\alpha_{\rm DW}$ that quantifies the gas accretion speed due to
disc winds and $\lambda$ that measures the angular removal efficiency of the
wind and hence dictates also the amount of gas lost to the wind). We will
therefore use the simpler $\alpha$-disc model in this paper, but we distinguish
between an ``accretion'' $\alpha$ that dictates the speed of the gas, and hence
the gas surface density for a given mass accretion rate onto the star, and a
``turbulent'' $\delta$ that measures the turbulent diffusion coefficient, and hence the scale height, of the pebbles. In this picture, the gas disc may still evolve rapidly under a high value of the viscous $\alpha$, even if the turbulence level $\delta$ is low.

\subsection{Disc temperature}

The temperature of the protoplanetary disc is given by a combination of stellar irradiation and viscous heating. For the former, we follow the temperature expression of \cite{Ida+etal2016} with the luminosity of the star following \cite{Johansen+etal2021},
\begin{equation}
  L(t) = 2 L_\odot \left( \frac{t}{\rm Myr} \right)^{-0.7} \exp\left[-\frac{1}{t/(0.15\,{\rm Myr})}\right] \, .
  \label{eq:Lsun}
\end{equation}
The first power law term, which is an approximate fit to the stellar evolution database of \cite{Baraffe+etal2015} for a solar-mass star, dominates after 0.5 Myr, while we added the exponential decline term to avoid excessive luminosities at earlier times. We compared our results to the luminosity model of \cite{Mori+etal2021}, which has a higher initial luminosity of the protostar, and found no significant difference to the resulting planet populations.

The viscous heating temperature, relevant in the inner regions of the disc for
the earliest evolution stages, is calculated from equation (B.10) in
\cite{Liu+etal2019a}, with the effective accretion rate that leads to heating
calculated as the full accretion rate multiplied by $\delta/\alpha$ to
distinguish between wind-driven angular momentum transport with parameter
$\alpha$ and turbulent heating with parameter $\delta$. Above $T=800\,{\rm K}$
we set $\delta=\alpha$ to reflect the onset of angular momentum transport by turbulence caused by the magnetorotational instability
\citep{DeschTurner2015}. The region of active MRI turbulence is nevertheless small
and short-lived and hence not very important for the outcome of our planet
formation calculations.

Adopting an evolving
evolving temperature and including viscous heating in the inner disc is not
formally consistent with the analytical evolution of the $\alpha$-disc, but
since the temperature evolution is generally small, we choose \citep[as
in][]{Liu+etal2019a} to use the temperature evaluated for a stellar luminosity $L_\star = L_\odot$ at the initial disc size
$R_1$ to obtain an approximation for the viscous evolution of the protoplanetary disc over its full life-time. We
additionally adopt the temporally declining analytical pebble flux from
\cite{Gurrutxaga+etal2024} when constructing the pebble column density in the
inner, viscously heated disc; the higher temperature in the inner disc will increase the
pebble speed and thereby decrease the local value of $Z$ slightly there compared
to the outer disc.

\subsection{Pebble flux}

\begin{figure}
  \centering
  \includegraphics[width=\linewidth]{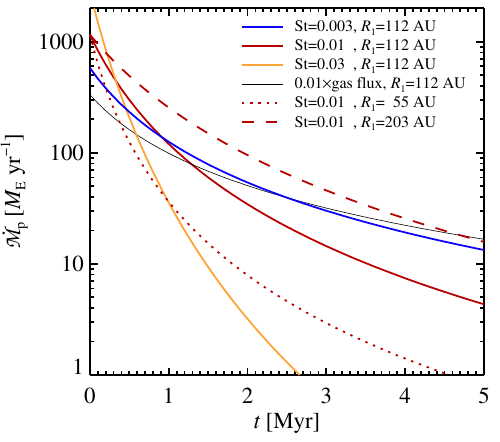}
  \caption{The pebble flux $\dot{\mathcal{M}}_{\rm p}$ through the inner disc as a function of time for
  three values of St (${\rm St}=0.003$, ${\rm St}=0.01$, ${\rm St=0.03}$) and three different initial protoplanetary disc sizes ($R_1=112\,{\rm AU}$, $R_1=55\,{\rm AU}$, $R_1=203\,{\rm AU}$) for the analytical pebble drift model of \cite{Gurrutxaga+etal2024} with constant ${\rm St}_\chi$ and $\alpha=0.01$. The thin black line indicates 1\% of the gas flux for the 112 AU disc; the ${\rm St}=0.003$ case follows gas depletion closely due to the limited radial drift. Larger pebbles drain out of the disc by radial drift on a shorter time-scale than the gas evolution time.}
  \label{f:pebbleflux_time}
\end{figure}

We show the pebble flux $\mathcal{\dot{M}}_{\rm p}$ through the inner regions of the protoplanetary disc as
a function of time in Figure \ref{f:pebbleflux_time} for three values of ${\rm
St}=0.003$, $0.01$, $0.03$, and three values of the initial disc size $R_1$. The three initial disc
sizes correspond to a starting stellar accretion rate of
$\dot{\mathcal{M}}_{\star0}=10^{-7}\,M_\odot\,{\rm yr}^{-1}$ that falls to
$\dot{\mathcal{M}}_{\star5}=5 \times 10^{-9}\,M_\odot\,{\rm yr}^{-1}$ (large disc), $\dot{\mathcal{M}}_{\star5} =2\times10^{-9}\,M_\odot\,{\rm yr}^{-1}$ (medium-sized disc) and $\dot{\mathcal{M}}_{\star5} = 10^{-8}\,M_\odot\,{\rm yr}^{-1}$ (very large disc) after
5 Myr of evolution. For the smallest value of ${\rm St}$, the pebble flux
follows approximately the constant metallicity expression
$\dot{\mathcal{M}}_{\rm p}(t) = Z_0 \dot{\mathcal{M}}_\star(t)$.  Pebbles with
larger ${\rm St}$ drain from the protoplanetary disc faster than the accretion
time-scale of the gas, decreasing the pebble metallicity with time. The
calculation presented in Figure \ref{f:pebbleflux_time} used, for simplicity, a
constant stellar luminosity of $L_\star = L_\odot$ and ignored viscous heating
in the inner disc, while the models presented in Sections \ref{s:pebble} and
\ref{s:map} will include viscous heating and
use the evolving luminosity from equation (\ref{eq:Lsun}).

\section{Conditions for the streaming instability}
\label{s:streaming}

A prominent behavior of the streaming instability is that dense, planetesimal-forming filaments only emerge in the sedimented mid-plane layer of pebbles when the average ``pebble metallicity'', defined as the ratio of the surface density of pebbles to the surface density of gas, is above a threshold value \citep{Johansen+etal2009b,BaiStone2010}, which is of order $\sim$1\% but depends on the Stokes number \citep{Carrera+etal2015,LiYoudin2021} as well as on the presence/strength of any background turbulence \citep{Johansen+etal2007,Yang+etal2018,SchaferJohansen2022,Lim+etal2024,Eriksson+etal2026}. We analyze here how the pebble metallicity in the inner regions of a protoplanetary disc depends on the properties of the disc and the pebble sizes. In Section \ref{s:map} we will show a map of how planetesimal formation by the streaming instability and planetary growth by pebble accretion depend on both the pebble Stokes number and on the strength of the turbulence in the disc. We use there the pebble metallicity threshold of \cite{LiYoudin2021}, which also agrees with simulations of weak turbulence ($\alpha$ between $10^{-4}$ and $10^{-3}$) driven by the magnetorotational instability \citep{Johansen+etal2007,Eriksson+etal2026}.

\subsection{Planetesimal formation at the water ice line}

The water ice line provides an excellent location to achieve a local metallicity increase, as there the pebbles slow down and pile up if their Stokes number changes upon losing the ice component \citep{IdaGuillot2016}. The released water vapor will also pile up interior of the ice line, with the possibility that the vapor diffuses outwards across the water ice line and increases both the sizes and the metallicity of the icy pebbles there \citep{RosJohansen2013,SchoonenbergOrmel2017,DrazkowskaAlibert2017}. In order to assess the potential metallicity increase at the water ice line without resorting to 3-D hydrodynamical simulations or evolving 1-D disc simulations with turbulent diffusion, we calculate the density of silicate pebbles and water vapor interior of the water ice line under the assumption that the flux of these components interior of the water ice line is equal to the pebble flux across the ice line. We add silicate pebbles and water vapor together as an estimate of the upper limit to the pebble enrichment that can be reached, since this combined metallicity is the highest achievable by diffusion of water vapor across the ice line. Also, water vapor interior of the water ice line will condense onto the silicate particles after the star undergoes a sudden decrease in its luminosity. \cite{RosJohansen2024} studied the effect of cooling after a stellar outburst and found that icy particles could grow up to ${\rm St}=0.03$ by ice condensation and sweep-up of silicate dust aggregates. Given the lack of experimental support for considering a lower fragmentation threshold for silicates than for water ice (see discussion in Section \ref{s:intro}), we consider silicate particles and combined silicate-ice particles to have the same Stokes number.
\begin{figure}
  \centering
  \includegraphics[width=\linewidth]{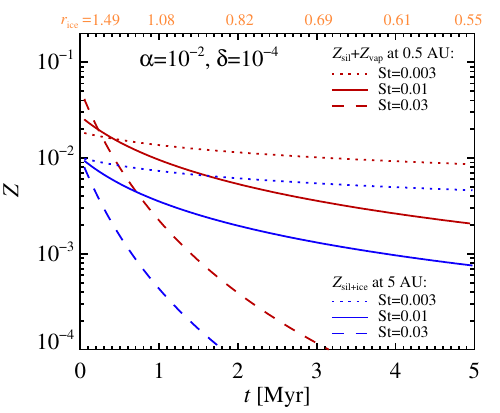}
  \includegraphics[width=\linewidth]{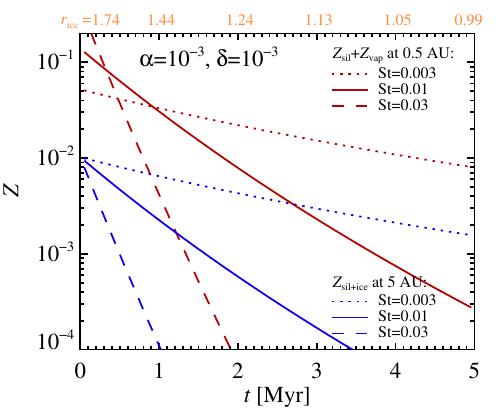}
  \caption{The panels show, for two different combinations of the viscosity
  coefficient $\alpha$ and the turbulent diffusion coefficient $\delta$, the
  temporal evolution of the pebble metallicity in a large protoplanetary
  disc ($R_1=112\,{\rm AU}$) at 5 AU (exterior of the water ice line, blue
  lines) and at 0.5 AU (interior of the water ice line, red lines). Water vapor released at the passage of the water ice line is included in the metallicity
  at 0.5 AU. The instantaneous position of the water ice line is indicated above the plots. The high viscous heating in the case $\alpha=\delta=10^{-3}$ (lower panel) pushes the ice line out, while
  the lower speed of the gas allows for a much bigger pile-up of water vapor
  interior of the water ice line.}
  \label{f:metallicity}
\end{figure}

\subsection{Temporal evolution of the metallicity}

In Figure \ref{f:metallicity} we show the temporal evolution of the metallicity
exterior (at 5 AU) and interior (at 0.5 AU) of the water ice line for three values of the global
Stokes number (${\rm St}=0.003$, ${\rm St}=0.01$ and
${\rm St}=0.03$) and two combinations of disc viscosity and turbulence ($\alpha=10^{-2}$ / $\delta=10^{-4}$ and $\alpha=10^{-3}$ / $\delta=10^{-3}$). We set the initial and final accretion rate to $\dot{\mathcal{M}}_{\star0}=10^{-7}\,M_\odot\,{\rm yr^{-1}}$ / $\dot{\mathcal{M}}_{\star5}=5 \times 10^{-9}\,M_\odot\,{\rm yr^{-1}}$ when $\alpha=10^{-2}$ and $\dot{\mathcal{M}}_{\star0}=10^{-8}\,M_\odot\,{\rm yr^{-1}}$ / $\dot{\mathcal{M}}_{\star5}=4.85 \times 10^{-9}\,M_\odot\,{\rm yr^{-1}}$ when $\alpha=10^{-3}$ in order to maintain the same disc size (112 AU) and initial disc mass ($\sim$500 $M_{\rm E}$ of solids) for the two values of $\alpha$. The metallicity interior of the water ice line includes, as discussed above, the
sublimated water vapor. The three blue lines show the evolution of the
metallicity as a function of time at a representative distance of 5 AU, based on
the analytical solution for the pebble flux from \cite{Gurrutxaga+etal2024}. The
global metallicity exterior of the water ice line starts at the initial
metallicity $Z_0=0.01$, but falls rapidly with time for ${\rm St}=0.01$ and
${\rm St}=0.03$, particularly for the model with $\alpha=10^{-3}$ due to the
much slower outwards gas speed in the outer disc. This viscous expansion of
gas and dust helps to store a large pebble reservoir in the outer disc, so that the pebble flux through the
inner disc is lower and longer-lasting \citep{Liu+etal2022}. In contrast, small pebbles
with ${\rm St}=0.003$ drift so slowly relative to the gas, for both the high and
low $\alpha$ case, that the pebble metallicity is maintained much better over the 5 Myr life-time of the protoplanetary disc.

The conditions interior of the water ice line are more conducive to planetesimal
formation than the outer regions, particularly if the Stokes number of the
pebbles exterior of the ice line is large, so that the pebble flux over the ice
line is high. If the water vapor would freeze out on the pebbles, as may happen
in the cooling phase after a stellar luminosity outburst
\citep{RosJohansen2024}, while maintaining the metallicity of 0.02--0.03, then
planetesimal formation could be triggered at the water ice line for turbulence
values $\delta$ between $10^{-5}$ and $10^{-4}$ (see Section \ref{s:map}). The
situation becomes better at lower $\alpha$ (lower panel of Figure
\ref{f:metallicity}): pile-ups are stronger due to slower gas speed. We will
nevertheless show in Section \ref{s:pebble} that models with low $\alpha$ in the inner disc and a
high $\dot{\mathcal{M}}_\star$ have very high planetary migration rates and hence struggle to
maintain planets in the 1 AU region.

\subsection{Formation of planetesimals beyond the water ice line}

Clearly, planetesimal formation cannot operate only around the water ice line
and is also required to function further out in the disc. Planetesimal formation
in the 10 AU -- 100 AU region may become easier at the higher-than-solar
metallicity reached by stars formed relatively recently in our Galaxy. The presence of pressure bumps in the outer disc regions, emerging perhaps from inverse cascade of magnetic energy in weak magnetorotational turbulence \citep{Lyra+etal2008a,Johansen+etal2009a,Eriksson+etal2026}, could facilitate the formation of planetesimals there. Also, the
CO and CO$_2$ ice lines are prime candidates for metallicity increases similar
to what occurs at the water ice line
\citep{Stammler+etal2017,SchneiderBitsch2021}. We will nevertheless not study the conditions for planetesimal formation around these colder ice lines in this paper, since that would require a more realistic model of pebble and vapor pile-ups in the outer disc.

\section{Planetesimal masses and growth time-scales}
\label{s:planetesimals}

We now turn to analyzing the characteristic mass of planetesimals that form by the streaming instability as well as their initial growth time-scale.

\subsection{Planetesimal  masses}

Most hydrodynamical simulations of the streaming instability are performed in the local shearing box frame where the natural length-scale is the gas scale-height $H$ and the natural density scale is the local mid-plane gas density $\rho_{\rm g}$ \citep{Johansen+etal2015,Simon+etal2016,LiYoudin2021}; the right-hand-side of the Poisson equation for the self-gravity of the pebbles ($\nabla^2 \varPhi = 4 \pi G \rho_{\rm p}$) can then be rewritten as the dust-to-gas ratio $\epsilon=\rho_{\rm p}/\rho_{\rm g}$ times the dimensionless self-gravity parameter
\begin{eqnarray}
  \varGamma &=& \frac{4\pi G\rho_{\rm g}}{\varOmega^2} = \frac{4 G \dot{\mathcal{M}}_\star}{3 \sqrt{2 \pi} \alpha c_{\rm s} H^2 \varOmega^2} = \frac{4 G \dot{\mathcal{M}}_\star}{3 \sqrt{2 \pi} \alpha c_{\rm s}^3} \nonumber \\
  && \approx 0.11 \left(\frac{\dot{\mathcal{M}}_\star}{10^{-7}\,M_\odot\,{\rm yr}^{-1}}\right) \left( \frac{\alpha}{10^{-2}} \right)^{-1} \left( \frac{T}{100\,{\rm K}} \right)^{-3/2} \, .
  \label{eq:Gamma}
\end{eqnarray}
The second equality relation is valid for the inner regions of an $\alpha$-disc where the gas density is related to the gas accretion rate through $\dot{\mathcal{M}}_\star = 3 \pi \nu \varSigma_{\rm g} = 3 \pi (\alpha c_{\rm s} H)(\sqrt{2\pi} H \rho_{\rm g})$. \cite{Liu+etal2020} analyzed a range of literature results on the masses of planetesimals formed by the streaming instability, published by different groups (using also independent codes), and found that the characteristic mass of planetesimals follows the approximate expression
\begin{equation}
  M_{\rm SI} = 5 \times 10^{-5} M_{\rm E} \left( \frac{\varGamma}{\pi^{-1}} \right)^{1.5} \left( \frac{H/r}{0.05} \right)^3 \left( \frac{M_\star}{M_\odot} \right) \, .
  \label{eq:MSI}
\end{equation}
We will fix the stellar mass to $M_\star=M_\odot$ from now on. 
\begin{figure*}
    \centering
    \includegraphics[width=0.49\linewidth]{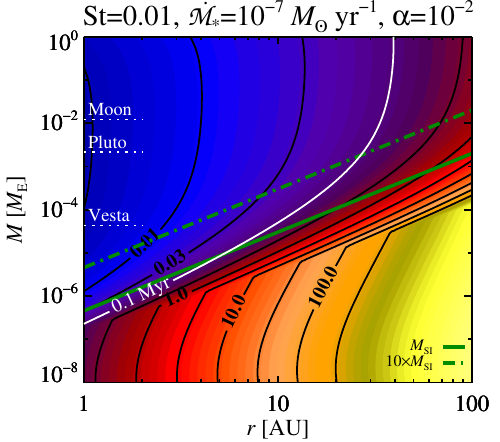}
    \includegraphics[width=0.49\linewidth]{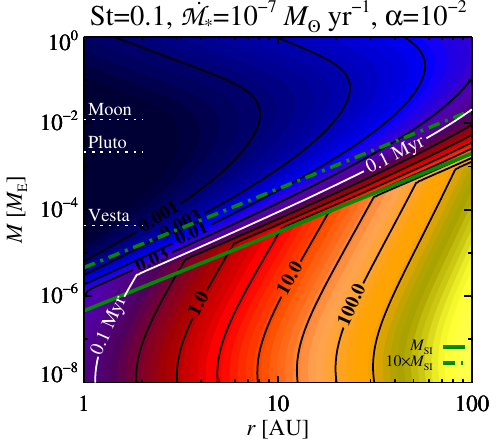}
    
    \caption{The two panels show the growth time-scale of planetesimals as a
    function of their distance to the star and their mass (colors and contours),
    overplotted with the characteristic mass of planetesimals formed by the
    streaming instability (green line) and 10 times the characteristic mass to
    represent the upper end of the initial mass function (green dash-dotted
    line). The left panel shows growth time-scales for ${\rm St}=0.01$ and the right panel for ${\rm St}=0.1$. We emphasize the 0.1 Myr contour line as a threshold time-scale that demarcates efficient planetary growth from inefficient growth at lower masses due to the transition the loosely coupled Bondi regime of pebble accretion.}
    \label{f:planetesimal_masses}
\end{figure*}

Using $c_{\rm s}\,$$\propto$$\,r^{-3/14}$ and $H/r\,$$\propto$$\,r^{2/7}$ from the temperature profile of
\cite{Ida+etal2016}, equation (\ref{eq:MSI}) reveals a radius scaling $M_{\rm SI} \propto r^{51/28}$ with the exponent
$51/28 \approx 1.82$. The full scaling of the planetesimal mass of equation
(\ref{eq:MSI}), with the temperature profile embedded, becomes 
\begin{eqnarray}
  M_{\rm SI} &=& 4.5 \times 10^{-7} M_{\rm E} \left( \frac{\dot{\mathcal{M}}_\star}{10^{-7}\,M_\odot\,{\rm yr}^{-1}} \right)^{1.5} \left( \frac{\alpha}{10^{-2}} \right)^{-1.5} \nonumber \\
  && \times \left( \frac{L_\star}{L_\odot} \right)^{-3/14} \left( \frac{r}{\rm AU} \right)^{51/28} \, .
  \label{eq:SI2}
\end{eqnarray}

\noindent We show in Figure \ref{f:planetesimal_masses} the characteristic planetesimal
mass formed by the streaming instability from 1 AU to 100 AU around a solar-mass
star at an early disc evolution stage where $\dot{\mathcal{M}}_\star=10^{-7}\,M_\odot\,{\rm
yr}^{-1}$.

\subsection{Growth time-scale}

We show in Figure \ref{f:planetesimal_masses} also the $e$-folding growth time-scale by pebble
accretion of a planetesimal or protoplanet of mass $M$ \citep[based on][]{JohansenLambrechts2017}, defined as $\tau_{\rm grow} = M/\dot{M}$, for
two different values of the Stokes number St. For simplicity, we use a
constant pebble Stokes number (${\rm St}=0.01$ or ${\rm St}=0.1$), constant turbulence strength ($\delta=10^{-4}$)
and constant pebble metallicity ($Z=0.01$) across the protoplanetary disc and
set the stellar luminosity equal to the current solar luminosity (ignoring the
early stellar luminosity evolution from equation \ref{eq:Lsun}).

Overall, Figure \ref{f:planetesimal_masses} demonstrates that growth time-scales
increase from $\sim$0.03 Myr to $\sim$3 Myr between 1 AU and 100 AU for ${\rm
St}=0.01$ at the characteristic mass of the streaming instability, and that the growth time-scales
are significantly longer at the streaming instability mass for the higher ${\rm St}=0.1$. However, for both values
of St, the growth time-scales drop dramatically at just ten times the
characteristic mass. Considering instead lower initial masses, Figure
\ref{f:planetesimal_masses} also implies that a putative generation of low-mass
planetesimals, similar to the objects in the classical Kuiper Belt, formed at
the late evolutionary stages of the protoplanetary disc with a low value of
$\varGamma$ \citep{Carrera+etal2017,Johansen+etal2025,LiChiang2025} would undergo
very little growth by pebble accretion.

The most
massive planetesimals formed by the streaming instability are indeed typically
at least 10 times more massive than the characteristic mass
\citep{Schafer+etal2017,Liu+etal2020,Schafer+etal2024}. We will therefore use 20
times the streaming instability mass as the nominal starting point for our pebble
accretion calculations in Section \ref{s:pebble} and Section \ref{s:map}, and we emphasize that the planetesimal starting mass is
very important, since the rather massive planetesimals formed by the streaming instability do not grow appreciably
by mutual collisions beyond a few AU in the protoplanetary disc \citep{LorekJohansen2022,Kaufmann+etal2025}.

\subsection{Pebble accretion threshold mass}

As demonstrated in \cite{Lyra+etal2023}, the inclusion of a size distribution of
pebbles can actually decrease the growth time-scale, particularly for large
values of ${\rm St}$. The reason is that there exists a planetesimal mass scale
below which the pebbles couple to the gas flow on longer time-scales than the
scattering process, which diminishes the kinetic energy dissipation and hence reduces the 
pebble accretion cross section significantly. This pebble accretion threshold mass $M_{\rm PA}$ is given by
the condition
\begin{equation}
  \frac{G M_{\rm PA}}{(\Delta v)^3} = \tau_{\rm f} \, ,
\end{equation}
where we equaled the time-scale to cross the Bondi radius, $t_{\rm B}=R_{\rm B}/\Delta v = G M / (\Delta v)^3$, to the friction time $\tau_{\rm f}$. The approach speed $\Delta v$ equals the sub-Keplerian speed of the pebbles if the planetesimal itself is low-mass and on a circular and non-inclined orbit. This yields a mass for the onset of pebble accretion of
\begin{equation}
  M_{\rm PA} = \frac{{\rm St}\, (\Delta v)^3}{G \varOmega} \approx 1.6 \times 10^{-6}\,M_{\rm E}\, \left( \frac{\rm St}{0.01} \right) \left( \frac{r}{\rm AU} \right)^{12/7} \, .
  \label{eq:MPA}
\end{equation}
We here did not include a numerical factor of 1/8 that is sometimes included in the expression for the loose coupling mass \citep{Ormel2017,Lyra+etal2023}.
Both the mass prefactor ($1.6 \times 10^{-6} M_{\rm E}$) and the distance
exponent $12/7 \approx 1.71$ are surprisingly close to the characteristic mass
of planetesimals formed by the streaming instability from equation
(\ref{eq:SI2}).

\subsection{Similarity between the streaming instability mass and the pebble accretion threshold mass}

The similarity between the loose coupling mass of pebble accretion and the characteristic mass of the streaming instability is actually not a coincidence. We expect that the mass of planetesimals formed by the streaming instability should broadly take the form
\begin{equation}
  M_{\rm SI} \sim \rho_{\rm R} (\eta r)^3 (\rho_{\rm g}/\rho_{\rm R})^p
  \label{eq:MSIc}
\end{equation}
Here, $\rho_{\rm R} = 9 \varOmega^2/(4 \pi G)$ is the Roche density, which must
be reached before a portion of a dense pebble filament can undergo full 3-D
collapse into a planetesimal. The characteristic scale of a filament formed by
the streaming instability is $\ell_{\rm SI} = \eta r$ \citep{YoudinGoodman2005},
where $\eta = -(1/2) (H/r)^2 \partial \ln P/\partial \ln r$ is a measure of the
radial pressure support of the gas, which feeds streaming motion between gas and
pebbles. The sub-Keplerian speed of the gas, in the absence of drag-force from
the pebbles, is given by $\Delta v = \eta v_{\rm K}$. However, the pebble
density in the mid-plane of the protoplanetary disc is similar to the mid-plane
gas density $\rho_{\rm g}$, which is orders of magnitude below the Roche
density. We therefore posit that the filament has to undergo first a radial
contraction before reaching the Roche density, shrinking initially only its radial extent by
a factor $\rho_g/\rho_{\rm R} = (1/9) \varGamma$. Following this radial contraction to reach the Roche density, further isotropic collapse from a
sphere of density $\rho_{\rm R}$ and length-scale $\ell_{\rm R} = \eta r
(\rho_g/\rho_{\rm R})$ corresponds to $p=3$ in equation (\ref{eq:MSIc}), while
smaller values of $p$ imply that the planetesimal collects mass from the larger
scale $\eta r$ in one or more directions. Computer simulations of the streaming
instability are consistent with $p=1.5$, as noted in the discussion of equation
(\ref{eq:MSI}) above. This choice of $p=1.5$ corresponds also to the value
suggested in \cite{KlahrSchreiber2021}, their equation (2), based on a model
of how self-gravity has to operate on sufficiently large scales to overcome
the turbulent diffusion generated by the streaming instability itself.

Using $\eta = \Delta v/v_{\rm K}$, equation (\ref{eq:MSIc}) can now be converted to
\begin{equation}
  M_{\rm SI} \sim \frac{9 \varOmega^2}{4 \pi G} \frac{\Delta v^3}{\varOmega^3} (\varGamma/9)^p \sim \frac{9^{1-p}}{4 \pi} \frac{\Delta v^3}{G \varOmega} \varGamma^p \, .
  \label{eq:MSI3}
\end{equation}
This expression has the same dependency on $\Delta v$ and $\varOmega$ as the
pebble accretion threshold mass from equation (\ref{eq:MPA}), with the ratio
\begin{equation}
  \frac{M_{\rm SI}}{M_{\rm PA}} = \frac{\varGamma^p}{\rm St} \frac{9^{1-p}}{4 \pi} \approx 0.084 \left( \frac{\rm St}{0.01} \right)^{-1} \left( \frac{\varGamma}{0.1} \right)^{1.5} \, .
\end{equation}
Here we took $p=1.5$ in the second equality. Hence, for a nominal Stokes number
(${\rm St}\sim0.01$) and a protoplanetary disc mass corresponding to the
earliest stages of disc evolution ($\varGamma\sim0.1$), the pebble accretion
threshold mass is approximately 12 times the characteristic mass of planetesimals formed by
the streaming instability; this order of magnitude difference is also clear from Figure \ref{f:planetesimal_masses}.

\subsection{Constant pressure support}

The scaling of the characteristic planetesimal mass from simulations, equation
(\ref{eq:MSI}), with $(H/r)^3$ rather than $(\Delta v)^3$, as in equation
(\ref{eq:MSI3}),  likely arises from the fact that the pressure support parameter $\varPi =
\Delta v/c_{\rm s}$ is typically kept constant in shearing sheet simulations.
This converts equation (\ref{eq:MSI3}) to the form
\begin{equation}
  M_{\rm SI} \sim \rho_{\rm R} \left( \frac{\Delta v}{c_{\rm s}} \right)^3 H^3 (\varGamma/9)^p \sim M_\star \varPi^3 \left( \frac{H}{r} \right)^3 (\varGamma/9)^p \, .
\end{equation}
The last equality came from expanding the Roche density to transform the $H^3$ scaling to an $(H/r)^3$ scaling similar to equation (\ref{eq:MSI}).

\section{Outcome of pebble and gas accretion}
\label{s:pebble}

We now investigate the growth of planetesimals by pebble accretion
and gas accretion starting at the upper range of the initial mass of planetesimals formed by the streaming instability.
\begin{figure*}
    \centering
    \includegraphics[width=0.40\linewidth]{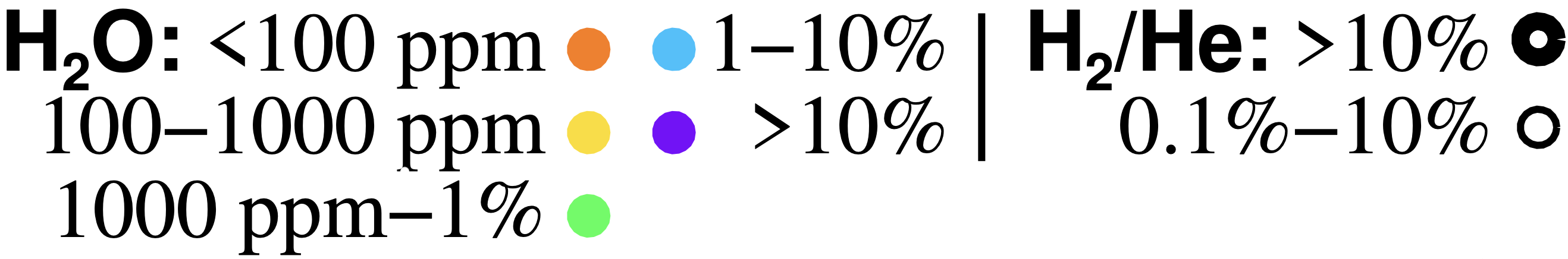} \\ \vspace{0.4cm}
    \includegraphics[width=0.45\linewidth]{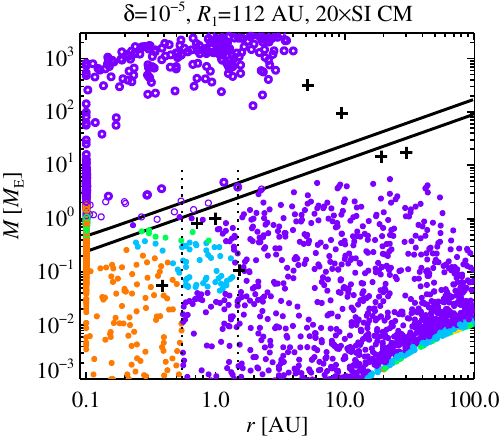}
    \includegraphics[width=0.45\linewidth]{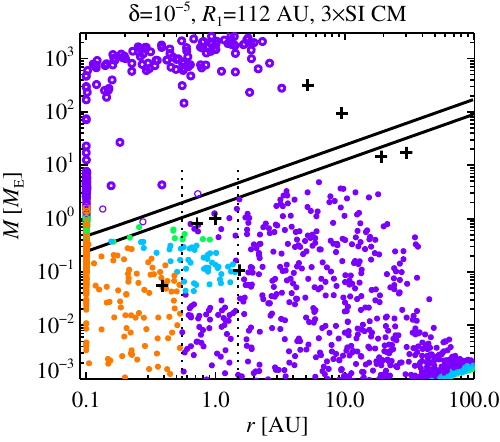} \\
    \includegraphics[width=0.45\linewidth]{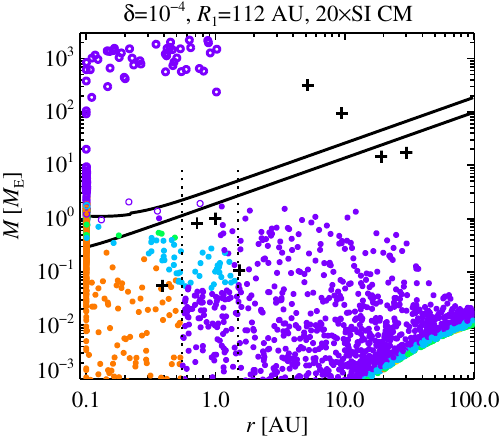}
    \includegraphics[width=0.45\linewidth]{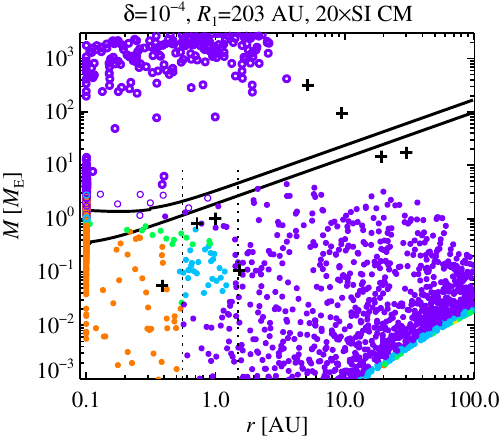}
    \caption{{\it Top left:} Population synthesis for a low turbulence level of
    $\delta=10^{-5}$ and an initial planetesimal mass that is 20 times
    the characteristic mass of the streaming instability (SI CM). Colors mark the
    water mass fraction in the non-gas material, while open circles indicate two gas fraction intervals (see
    legend on top). The two black lines show the pebble isolation mass at 
    early ($t=0.3$ Myr) and late ($t=5.0$ Myr) stages in the evolution of the disc, while the
    vertical dotted lines indicate the early and late locations of the water ice
    line. The Solar System planets are indicated with plus symbols. We see the
    emergence of all classes of planets, including cold gas giants akin to
    Jupiter and Saturn. {\it Top right:} Result of decreasing the planetesimal
    masses to 3 times the characteristic mass. This has mainly the effect of reducing the
    planetesimal growth rates beyond 20 AU, due to the transition there towards the loose coupling branch of pebble accretion, which decreases the population of warm and cold gas
    giants. {\it Bottom left:} Increasing the turbulence strength instead to
    $\delta=10^{-4}$ suppresses the formation of cold gas giants. {\it Bottom
    right:} Cold gas giants can form again with $\delta = 10^{-4}$ turbulence when 
    the initial disc size is increased to $R_1=203\,{\rm AU}$.}
    \label{f:delta}
\end{figure*}

\subsection{Model assumptions}

Our algorithms used for calculating the pebble accretion and
gas accretion rates, as well as the planetary migration rate and gap formation,
follow \cite{JohansenLambrechts2017}, \cite{Ida+etal2018} and
\cite{Johansen+etal2019}. The pebble accretion algorithm includes both the 3-D and 2-D phases of the Bondi and Hill accretion regimes, with a smooth transition between them based on a calculation of the accretion impact parameter \citep{JohansenLambrechts2017}. We ignore positive torques from the pebble flow close the planet, which can slow down or even reverse migration at Stokes numbers above ${\rm St}=0.03$ \citep{Benitez-LlambayPessah2018}. We also ignore positive torques arising from viscous heating in the inner disc \citep{Paardekooper+etal2010,Paardekooper+etal2011,Bitsch+etal2015}, from the planetary accretion heat \citep{Benitez-Llambay+etal2015,BaumannBitsch2020} as well as the torque reduction from the dynamical corotation torque \citep{Paardekooper2014}. The latter has been demonstrated to potentially slow down Type I migration by a factor of $\sim$2 \citep{Ndugu+etal2021,Yun+etal2022}, which motivates our exploration of models with a lower gas fraction (and hence slower migration) in Section \ref{s:lowgas}. By not including multiple potential sources of positive torques from the gas, our study thus constitutes a ``worst case scenario'' considering planetary migration at the unabated Type I and Type II rates. We discuss this further in Section \ref{s:summary}.

Gas accretion is
allowed either when the mass of a protoplanet is higher than the local pebble
isolation mass, whose dependence on temperature and viscosity is based here on
\cite{Bitsch+etal2018}, or when the pebble accretion time-scale is longer than
10 Myr \citep{Gurrutxaga+etal2024}, which is often the case late in the evolution of
the disc as the pebble flux diminishes. We set the opacity level in the envelope to $\kappa_{\rm env}=0.005\,{\rm m^2\,kg^{-1}}$. A higher or lower opacity will slow down or speed up the Kelvin-Helmholtz gas contraction phase of the envelope \citep{Ida+etal2018,Johansen+etal2019}, but we checked that even factor ten changes to the opacity will only change the final orbits of cold gas giants in our simulations by $\sim$1 AU.

We use a nominal pebble metallicity of
$Z=0.01$, consisting of 50\% silicate/metal and 50\% water ice
\citep{Lodders2003}. The water ice fraction is removed from the pebbles when
they reach the temperature of $T_{\rm ice}=170\,{\rm K}$ in either the
protoplanetary disc or at the radiative-convective boundary (RCB) in the gas
envelopes of massive protoplanets. The temperature at the RCB is calculated as a
function of the local disc properties, planet mass and accretion rate from the
analytical expressions of \cite{PisoYoudin2014}; our implementation is detailed
in \cite{Johansen+etal2023c}.

We assume that all pebbles have the same product of Stokes number and
logarithmic pressure gradient, with a nominal value of ${\rm St}=0.01$ at the
characteristic disc radius $R_{\rm c}$. The slightly shallower pressure gradient in the main disc regions, interior of the characteristic radius $R_{\rm c}(t)$ (see equation \ref{eq:Rc}), yields then a Stokes number there that is approximately 30\%
higher than at $R_{\rm c}$. We ignore the presence of smaller dust grains in the size
distribution and refer to \cite{Lyra+etal2023} for a discussion of the role of the small end of the size distribution for pebble accretion. We do not consider the presence of any local pressure bumps in our growth track calculations \citep{JiangOrmel2023,Eriksson+etal2026}.

We also assume that the growing planets have near-zero eccentricity and near-zero
inclination, as is indeed observed for the most massive bodies that grow by
pebble accretion in $N$-body simulations
\citep{Levison+etal2015,Liu+etal2019b,Yzer+etal2025,LorekLambrechts2026}. The most important
simplification of our work is therefore arguably that we ignore any interactions
between the growing planets, which means that we cannot capture protoplanet
collisions nor eccentricity and inclination excitation or trapping in resonant
chains \citep{Lambrechts+etal2019,Bitsch+etal2019,Izidoro+etal2021}. However,
this simplification allows us to create population plots whose statistical properties
can easily be compared to the Solar System and to exoplanets at the population
level, although it gives no information about the range of architectures of
actual planetary systems.

Our nominal protoplanetary disc model has viscous $\alpha=10^{-2}$, an initial
gas accretion rate of $\dot{\mathcal{M}}_{\star0}=10^{-7}\,M_\odot\,{\rm yr}^{-1}$ and a final
accretion rate of $\dot{\mathcal{M}}_{\star5}=5 \times 10^{-9}\,M_\odot\,{\rm yr}^{-1}$ after 5 Myr of
evolution. The initial size of such a disc is $R_1=112$ AU, representing a disc
among the 10\% largest observed at the class 0 and I phases
\citep{NajitaBergin2018,Bate2018,Ohashi+etal2023,TobinSheehan2024}. Gas giant planets are
quite rare at solar metallicity -- only around 7\% of solar-metallicity stars
host a gas giant with a mass higher than Saturn's in the 1--5 AU region
\citep{Fulton+etal2021} -- hence we probe the largest (and thus
rarest) protoplanetary
discs for their formation. We also experiment with even larger discs (with final
accretion rate $\dot{\mathcal{M}}_{\star5}=10^{-8}\,M_\odot\,{\rm yr}^{-1}$ and a size
of 203 AU) as well as more nominal-sized discs (with final accretion rates
$\dot{\mathcal{M}}_{\star5}=2 \times 10^{-9}\,M_\odot\,{\rm yr}^{-1}$ / $\dot{\mathcal{M}}_{\star5}=
10^{-9}\,M_\odot\,{\rm yr}^{-1}$ and initial sizes 55 / 33 AU). The latter are
more prone to forming super-Earths and sub-Neptunes as their upper mass range,
rather than gas giants.
\begin{figure}
    \centering
    \includegraphics[width=0.9\linewidth]{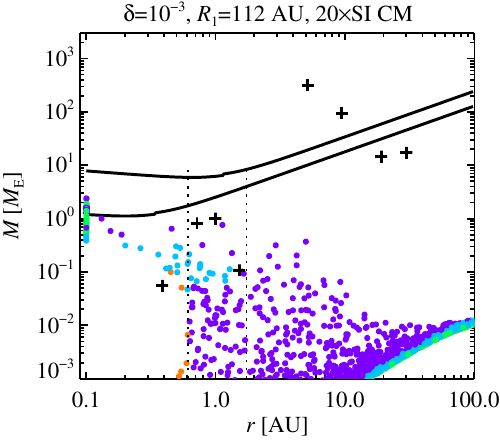} \\
    \includegraphics[width=0.9\linewidth]{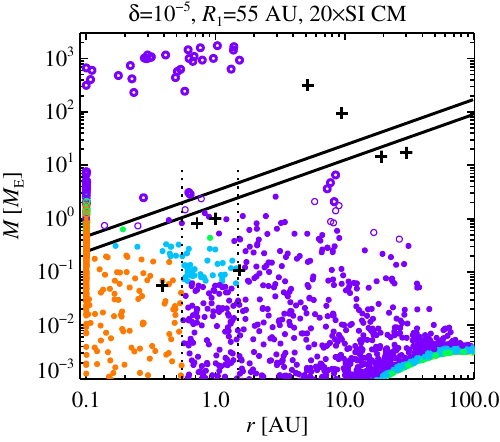} \\
    \includegraphics[width=0.9\linewidth]{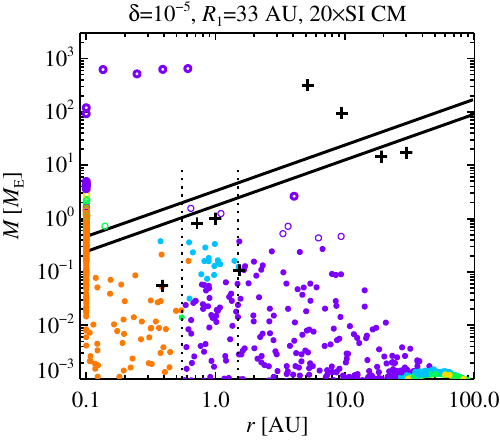} \\

    \caption{The plots illustrate how increasing the turbulence level further than in Figure \ref{f:delta} (to $\delta=10^{-3}$, top panel) or lowering
    the initial disc size (to $R_1=55$ AU, middle panel, or $R_1=33\,{\rm AU}$,
    bottom panel) affects the ability of the disc to form planets via pebble accretion.
    Increasing the turbulence level also increases the temperature in the inner
    disc, which manifests in the top plot as a higher pebble isolation mass in
    the inner disc and a reduction of the pebble accretion rate there.
    The smaller discs with weaker turbulence, in contrast,  retain their
    ability to form rocky planets, super-Earths and sub-Neptunes.}
    \label{f:nogasgiants}
\end{figure}

The turbulence level $\delta$ controls both the pebble scale-height and the turbulent heating of the protoplanetary disc. We explore turbulence strengths between $\delta=10^{-5}$ and $\delta=10^{-3}$. The lower and middle range of this interval is supported by observations of the scale-height of dust in two well-resolved edge-on protoplanetary discs \citep{Pinte+etal2016,Villenave+etal2022}, while the higher range represents conditions in a few well-resolved discs with high pebble scale-heights as well as the typical upper values from many discs with unresolved pebble scale-heights \citep{Villenave+etal2025}.

\subsection{Turbulence strength, planetesimal masses and disc size}

Figure \ref{f:delta} shows the resulting planet populations for two different
values of $\delta$, namely $\delta = 10^{-5}$ (top panels) and
$\delta=10^{-4}$ (lower panels). For the lower value of $\delta$, we observe the
formation of rocky planets, super-Earths, sub-Neptunes as well as hot and cold
gas giants when using a starting planetesimal mass that is 20 times the
characteristic mass from equation (\ref{eq:MSI}). The disc loses its cold gas giants in the 5 AU region when we start instead at 3 times the characteristic mass
(top right panel of Figure \ref{f:delta}), highlighting the  importance
of the planetesimal starting mass and confirming that the planetesimal growth
rates shown in Figure \ref{f:planetesimal_masses} strongly affect the final
planetary system. For the higher value of $\delta=10^{-4}$ (bottom panels of
Figure \ref{f:delta}), the disc also reduces the orbits of its gas giants, but the lower right plot shows that doubling the
initial disc size to 203 AU reinstates Jupiter and Saturn analogues under such turbulent
conditions.

The fact that cold gas giants require favorable conditions to form is further
illustrated in Figure \ref{f:nogasgiants}. The top panel shows that increasing
the strength of the turbulence to $\delta=10^{-3}$, while maintaining a viscosity of $\alpha=10^{-2}$, quenches the formation of all
classes of gas giants: cold, warm and hot. The increased turbulence decreases
the growth rates of protoplanets both by increasing the scale-height of the
pebble mid-plane layer, which means that protoplanets spend more time in the 3-D branch of pebble accretion, and by increasing the temperature in the inner disc by viscous heating, which decreases the Bondi radius for pebble accretion \citep{BatyginMorbidelli2022,Danti+etal2025}. The increased temperature also increases the pebble isolation mass in the inner disc and makes it harder to form sub-Neptunes with small gas envelopes.

In the middle and bottom panels of Figure \ref{f:nogasgiants}, we instead
decrease the initial disc size from $R_1=112$ AU to more median values of
$R_1=55$ AU and $R_1=33$ AU. This removes first cold gas giants and, for the
smallest initial size, even hot and warm gas giant become very rare. The colder conditions in the
inner disc, compared to the case of increased turbulence strength, nevertheless
still allow super-Earths and sub-Neptunes to form and migrate to hot orbits.
\begin{figure}
    \includegraphics[width=0.9\linewidth]{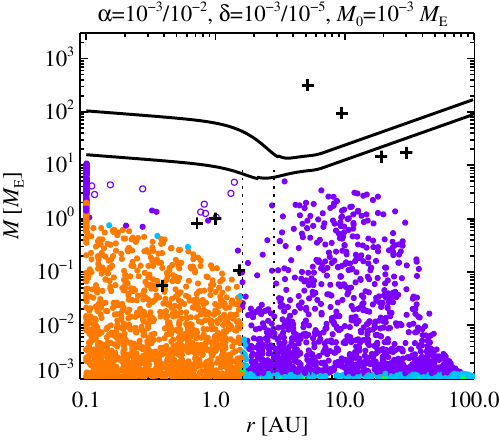}
    \includegraphics[width=0.9\linewidth]{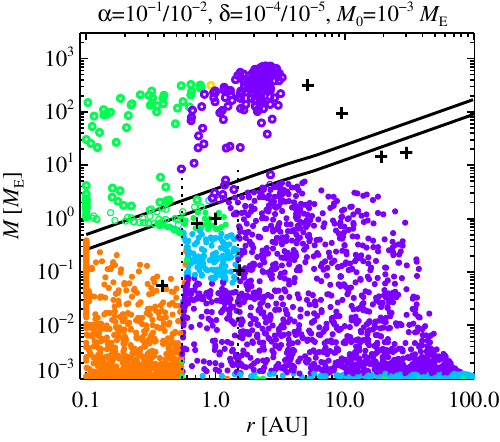}
    \caption{Comparison of two models with angular momentum transport
    coefficient $\alpha=10^{-2}$ exterior of 5 AU and a lower value
    ($\alpha=10^{-3}$, top panel) or a higher value ($\alpha=10^{-1}$, top
    panel) interior of 5 AU. We fix the initial planetesimal mass to $M_0=10^{-3}\,M_{\rm E}$ and starting times between 0.5 and 5 Myr, to probe the effect of early collisional evolution. In the low-$\alpha$ case, we use
    $\delta=\alpha=10^{-3}$ to additionally probe the effect of viscous heating, which pushes
    the ice line outwards to 1.5--3 AU. Pebble accretion is strongly suppressed
    by the high temperatures in the inner disc and all planets ending up in this
    region have migrated from more distant orbits beyond the water ice line. In the
    case of higher $\alpha$, migration is strongly reduced due to lower gas
    column densities in the inner disc, which suppresses the migration of
    super-Earths and sub-Neptunes to the disc edge. Lower-mass rocky planets
    form more or less in-situ with this reduced migration.}
    \label{f:twoalpha}
\end{figure}
\begin{figure}
    \includegraphics[width=0.9\linewidth]{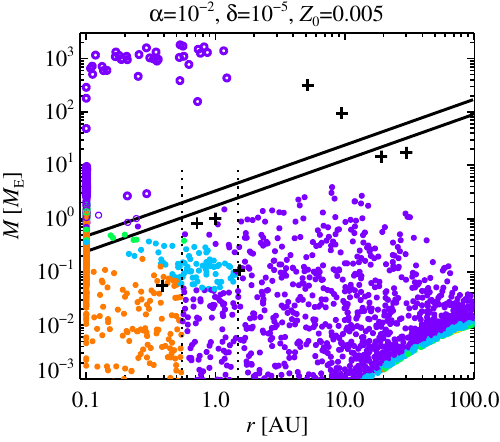}
    \includegraphics[width=0.9\linewidth]{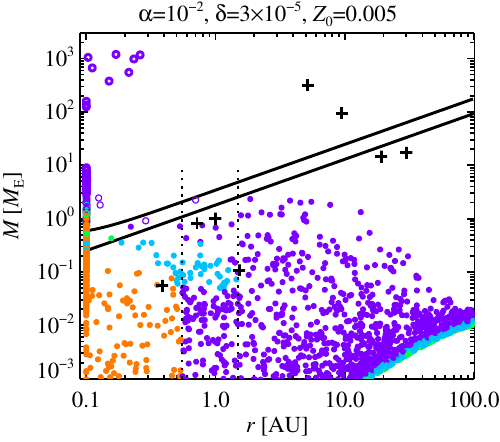}

    \caption{The two plots show the effect of halving the pebble metallicity to $Z=0.005$. {\it Top panel:} Maintaining the turbulence at a low level of $\delta= 10^{-5}$, the same as in the top left panel of Figure \ref{f:delta}, effectively quenches the formation of cold gas giants, while warm and hot gas giants can still form. The populations of sub-Neptunes, super-Earths and rocky planets are relatively unaffected by the lower metallicity. {\it Bottom panel:} Increasing the turbulence strength  to $\delta=3 \times 10^{-5}$ also quenches the formation of warm gas giants, with only a small population remaining in the 0.1 -- 0.3 AU region.}
    \label{f:lowZ}
\end{figure}

\subsection{Water mass fraction}

The water mass fraction is indicated with colors in the population plots. The water ice line (vertical dotted lines) starts at approximately 1.5 AU in the cold disc models of Figure \ref{f:delta} and migrates to nearly 0.5 AU after 5 Myr of evolution, due mainly to the steadily decreasing stellar luminosity. All giant planets in Figure \ref{f:delta} have very water-rich cores that have been assembled well exterior of the water ice line. The super-Earths that migrate to the inner disc edge without accreting gas are all very dry, while the sub-Neptunes in hot orbits mainly have water-rich cores \citep{Nielsen+etal2025}. In the 1 AU region, the water mass fraction decreases with increasing planet mass due to sublimation of ice from the accreted pebbles \citep{Johansen+etal2021,Wang+etal2023,Johansen+etal2023c}. Earth likely contains a few thousand parts per million (ppm) of water, when including the dominant reservoir in the core, which corresponds roughly to the green color in Figure \ref{f:delta}. At higher turbulence levels (lower right panel of Figure \ref{f:delta}), migration becomes more pronounced, due to lower pebble accretion rates, exposing the 1 AU region to water-rich (blue) interlopers that formed at so cold conditions that water vapor sublimated from the pebbles is released below the RCB and hence can not escape from the growing planet \citep{Wang+etal2023}. 

\subsection{Reducing or increasing $\alpha$}

We continue to experiment with a lower value and a higher value of the angular
momentum transport coefficient $\alpha$. Figure \ref{f:twoalpha} shows results
with $\alpha=10^{-3}$ and $\alpha=10^{-1}$ interior of 5 AU, while we fix
$\alpha=10^{-2}$ in the outer disc in order to maintain the same pebble flux as
in the top-left panel of Figure \ref{f:delta}. We additionally choose to fix the initial planetesimal mass here at $M_0=10^{-3}\,M_{\rm E}$. This is done partially to prevent the value of $\alpha$ in the inner disc from affecting the planetesimal masses, through its effect on the self-gravity parameter $\varGamma$ in equation (\ref{eq:Gamma}), which would complicate the comparison between low and high $\alpha$; and partially with the motivation to probe the effect of early collisional evolution within the protoplanetary disc \citep{LorekJohansen2022}. For the lower value of $\alpha$
(top panel of Figure \ref{f:twoalpha}), we also set the strength of the
turbulence to $\delta=\alpha=10^{-3}$ in order to mimic an inner disc where
angular momentum is transported by turbulence rather than by disc winds. The
combination of higher gas column densities and stronger turbulence leads to
significant viscous heating of the inner disc. The ice line is pushed out to
1.5--3 AU.  The top panel of Figure \ref{f:twoalpha} shows that pebble accretion
is efficiently suppressed by the high temperatures in the inner regions of the
disc \citep{BatyginMorbidelli2022,Danti+etal2025}. Planets ending in orbits interior of 2 AU have all migrated there from more distant, and hence cooler and more water-rich, starting orbits. The bottom panel of Figure
\ref{f:twoalpha} shows the effect of instead increasing the turbulent angular
transport coefficient to $\alpha=10^{-1}$. Such a fast gas accretion speed could
correspond to angular momentum transport by strong discs winds facilitated by
the Hall MHD effect in the inner regions of disc \citep{Mori+etal2019}; we
therefore set here a weaker turbulence level of $\delta=10^{-4}$.
Planetary migration is reduced due to the low gas column densities, and
sub-Neptunes and super-Earths experience diminished migration, while smaller rocky
planets do not migrate substantially from their starting point in such high-$\alpha$ conditions.
\begin{figure}
    \includegraphics[width=0.9\linewidth]{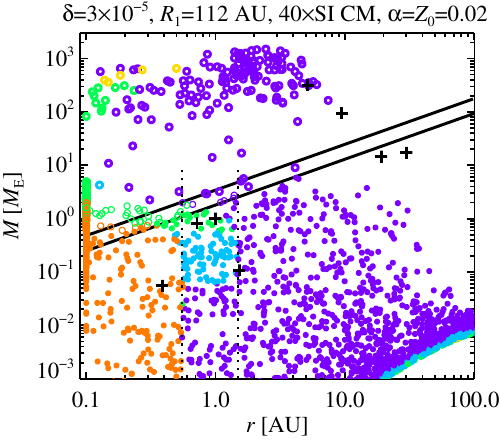}
    \includegraphics[width=0.9\linewidth]{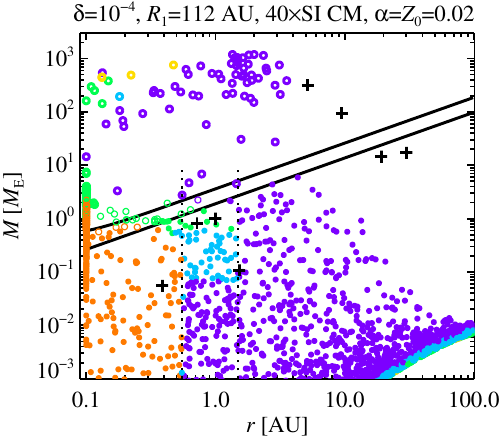}
    \caption{Two models with high $\alpha=0.02$ and high $Z=0.02$, giving the
    same dust mass as in the nominal model but with half the gas mass. This may
    represent either an initially higher metallicity or a gradual gas loss to disc winds. The lower migration rates
    make it possible to form Jupiter analogues both at $\delta=3 \times 10^{-5}
    $ (upper panel) and at $\delta=10^{-4}$ (lower panel).}
    \label{f:highZ}
\end{figure}

\subsection{Lower metallicity}

We now turn to exploring the effect of lowering the pebble metallicity to
$Z=0.005$. Figure \ref{f:lowZ} shows results for two values of the turbulence
strength. At $\delta=10^{-5}$, cold gas giants vanish,
while warm and hot gas giants still form if they start their growth very early in the
evolution of the disc. Increasing the turbulence level to $\delta=3 \times 10^{-5}$
(lower panel of Figure \ref{f:lowZ}) removes also most of the the warm gas giants. The
populations of sub-Neptunes, super-Earths and rocky planets are nevertheless
relatively unaffected by the lower metallicity
\citep{Liu+etal2019a,Nielsen+etal2023}, which agrees well with observations of
little or no metallicity-dependence for non-giant exoplanets
\citep{Buchhave+etal2012}.

\subsection{Lower gas fraction}
\label{s:lowgas}

The planetary populations produced under various conditions in the
protoplanetary disc in this section show as a common feature that forming cold
gas giants by pebble accretion requires low turbulence and/or large initial disc
sizes, while warm and temperate gas giants in $\sim$0.1--1 AU orbits emerge readily even under conditions where cold gas giant formation is suppressed, highlighting migration as a key element in making formation of cold gas giants difficult. We go on to test the effect of the migration rate for the formation of gas giant planets. We  maintain the disc size and total dust mass but half the amount of gas in the disc, as could be the consequence of gas loss due to disc winds \citep{Tabone+etal2022,ZhaoMatsumura2025}. We mimic the effect of gas loss here by increasing the viscosity to $\alpha=2 \times 10^{-2}$ while doubling the metallicity to $Z=0.02$. We maintain the approximate initial planetesimal mass, despite the lower gas mass and hence lower $\varGamma$, by starting here at 40 times the characteristic mass formed by the streaming instability. Figure \ref{f:highZ} shows the resulting populations for two values of the turbulence strength, $\delta=3 \times 10^{-5}$ and $\delta=10^{-4}$. Cold gas giants appear under both turbulence levels, which highlights the positive effect of gas removal for parking migrating planets in cold orbits. This overall demonstrates that even with rapid pebble accretion in the outer disc, reducing the migration rate of protoplanets leads to a better agreement with the population of cold gas giants.

\section{Mapping the conditions for streaming instability and pebble accretion}
\label{s:map}

We finally in this section map the conditions in the protoplanetary disc that allow planetesimal formation by the streaming instability in terms of Stokes number ${\rm St}$ and turbulence strength $\delta$, as well as the outcome of pebble accretion calculations starting from 20 times the characteristic mass of planetesimals formed by the streaming instability.
\begin{figure*}
    \centering
    \includegraphics[width=0.95\linewidth]{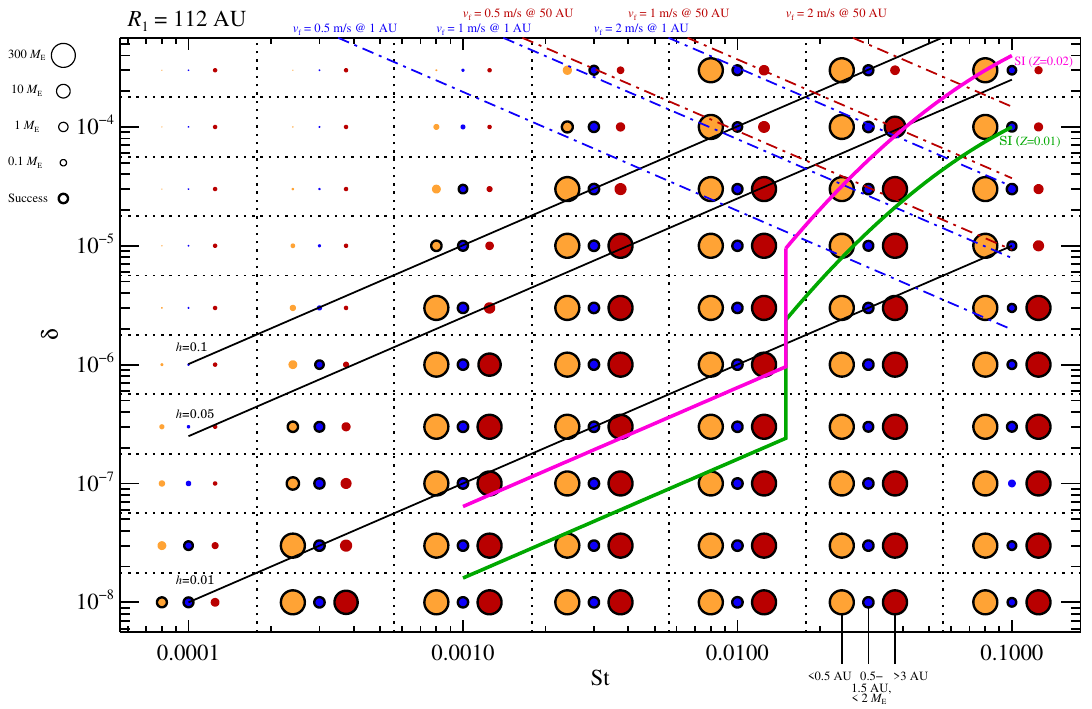}
    \includegraphics[width=0.95\linewidth]{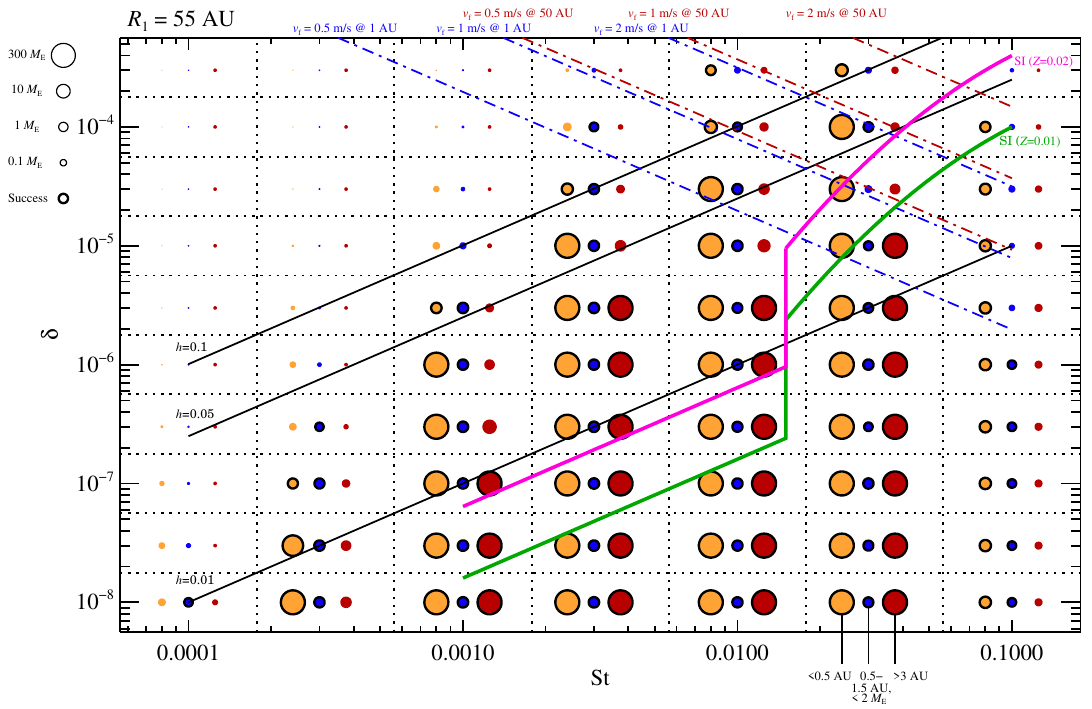}
    \caption{Map of conditions for triggering planetesimal formation by the streaming instability as well as the outcome of planet formation simulations, both as a function of Stokes number
    ${\rm St}$ and turbulent diffusion coefficient $\delta$. We use here a viscosity parameter for angular momentum transport of
    $\alpha=10^{-2}$ and an initial disc of either 112 AU (top) or 55 AU
    (bottom). We refer to the main text for a full discussion of these two plots.}
    \label{f:map}
\end{figure*}

\subsection{Streaming instability and turbulence}

We base the criterion for filament formation by the streaming instability on the threshold derived in \cite{LiYoudin2021} in terms of the mid-plane dust-to-gas ratio $\epsilon_0$. This dust-to-gas ratio is then converted to a critical turbulence strength $\delta_{\rm c}$, above which the critical mid-plane density cannot be reached due to turbulence inhibiting pebble sedimentation, by inverting the equation
\begin{equation}
  \epsilon_0 = Z \sqrt{\frac{{\rm St}+\delta_{\rm c}}{\delta_{\rm c}}} \, ,
  \label{eq:SIcrit}
\end{equation}
to yield $\delta_{\rm c}({\rm St},\epsilon_0({\rm St}))$ for a given Stokes number. Here, $Z$ is the ratio of the surface density of pebbles relative to the surface
density of the gas (i.e., the metallicity). We take this approach rather than
basing the metallicity threshold on simulations with forced turbulence  \citep{Lim+etal2024}, since simulations of weak turbulence caused by the magnetorotational instability show a mid-plane density threshold that is very similar to the non-turbulent case \citep{Eriksson+etal2026}. We also ignore the possibility that
turbulence excited by a physical instability could produce pressure
bumps as a side effect, which can aid the streaming instability by concentrating
pebbles towards the pressure maximum
\citep{Johansen+etal2007,Johansen+etal2009a,Yang+etal2018,Zhao+etal2025,Eriksson+etal2026}.

The vertical shear instability, which is a plausible source of turbulence in protoplanetary discs \citep{Nelson+etal2013}, has been demonstrated to allow the co-existence of a thin sedimented mid-plane layer of pebbles, where the streaming instability drives weak turbulence, together with significant large-scale vertical motion of the gas above and below the mid-plane \citep{SchaferJohansen2022,HuangBai2025}, effectively decoupling the weak turbulence experienced by the pebbles in the mid-plane from the stronger turbulence experienced by bulk of the gas. Even when the vertical shear instability is given a head start to evolve before the streaming instability, the vertical shear instability exerts a scale-dependent diffusion on the dust, with the diffusion coefficient over the thickness of the pebble layer measured at the level of $\delta \sim 7 \times 10^{-5}$ \citep{Schafer+etal2025}. The sedimented pebbles are  embedded in a large-scale up-down flow with a measured diffusion strength of $\delta \sim 4 \times 10^{-4}$ for pebbles with ${\rm St} = 0.03$ and $Z=0.02$ \citep{Schafer+etal2025} as well as for pebbles with ${\rm St} = 0.1$ and $Z=0.01$ \citep{HuangBai2025}. Pebbles in such a flow experience internal settling and low collision speeds driven by weak turbulence, while the pebble layer is forced to undulate around the mid-plane to a significant height dictated by the stronger large-scale turbulence (although the mid-plane layer height is significantly reduced by the drag-force exerted from the pebbles on the gas). The small-scale turbulent diffusion will thus dictate the fragmentation-limited growth of pebbles to proceed with a low value of $\delta$, while the large-scale turbulent diffusion will dictate the pebble accretion rate through its determination over the mean mid-plane pebble density. The simulations of \cite{Schafer+etal2025} and \cite{HuangBai2025} were nevertheless isothermal, which maximizes the strength of the vertical shear instability, and the effect of pebble back-reaction on the weaker turbulence that arises in simulations of the vertical shear instability with realistic cooling times \citep{Pfeil+etal2023} has currently not been probed.

We show the threshold metallicity for forming planetesimals by the streaming instability in
Figure \ref{f:map}, as a function of Stokes number ranging from ${\rm St}=
10^{-4}$ to ${\rm St}=10^{-1}$, for two values of the
metallicity (eq.~\ref{eq:SIcrit}): $Z=0.01$ and $Z=0.02$ (the latter reflecting gas removal by disc winds or a local pile-up of
pebbles e.g.\ at the H$_2$O or CO ice lines, as discussed in Section
\ref{s:streaming}). We overplot in Figure \ref{f:map} curves of constant relative mid-plane layer
thickness $h=H_{\rm p}/H$, where $H_{\rm p}$ is the pebble scale-height and $H$ is the gas scale-height, of $h=0.01$, $0.05$ and $0.1$, due the
direct observability of this physical quantity in protoplanetary discs
\citep{Pinte+etal2016,Villenave+etal2022}, given knowledge of the gas scale-height from thermal modeling of observations of small dust grains in scattered light. We also indicate, with dot-dashed
lines, the turbulent diffusion strength $\delta$ corresponding to a given Stokes
number ${\rm St}$ for three values of the pebble fragmentation threshold speed,
from equation (\ref{eq:fraglim}), at either 1 AU (blue) or 50 AU (red).

The abrupt jump in the $\delta$ threshold
around ${\rm St}\sim0.01$ in Figure \ref{f:map} is due to a transition in the
metallicity threshold from less than unity to $\epsilon_0 \approx 2.5$ when
decreasing the Stokes number below ${\rm St}=0.01$
\citep{Carrera+etal2015,Yang+etal2017,LiYoudin2021}; we note that this jump is not observed in simulations with imposed turbulence where the mid-plane density threshold at high ${\rm St}$ is significantly higher \citep{Lim+etal2024}; however those simulations with imposed background turbulence only probed turbulence with strength down to $\delta=10^{-4}$ and may not fully represent turbulence driven by a physical instability such as the vertical shear instability \citep{Schafer+etal2025} or the magnetorotational instability \citep{Eriksson+etal2026}.

\subsection{Planetary systems}

For three discrete values in each decade of ${\rm St}$ and $\delta$, we also perform
pebble accretion simulations starting from 20 times the characteristic mass of
planetesimals formed by the streaming instability. We quantify the outcome of
these simulations in terms of three symbolic circles in Figure \ref{f:map}:
\begin{enumerate}
  \item The yellow (left) circle marks the maximum mass of planets ending in
  orbits closer than 0.5 AU from the star;
  \item The blue (middle) circle marks the maximum mass of planets ending in
  orbits between 0.5 and 1.5 AU, with a cap at 2 $M_{\rm E}$ to avoid temperate
  gas giant interlopers;
  \item The red (right) circle marks the maximum mass of planets ending in
  orbits more distant than 3 AU.
\end{enumerate}
The three circles thus represent (1) planets in hot and warm orbits (hot/warm gas giants, hot/warm
sub-Neptunes and hot/warm super-Earths), (2) rocky planets in temperate Earth-like
orbits, and (3) planets in cold orbits. For each category, we define a success
criterion and mark the successful cases with a thick circle edge. For the yellow
class, the success criterion is to form super-Earths and/or sub-Neptunes with the threshold mass set at $M=1\,M_{\rm E}$, which we
take as the dividing line between rocky planets, of lower masses, and
super-Earths and sub-Neptunes. For the blue class, we take as
success criterion to reach a mass of at least $M=0.4\,M_{\rm E}$, which demarcates
significant growth of rocky planets by pebble accretion. Finally, for the red
class we adopt a success criterion of reaching a mass of $M=100\,M_{\rm E}$, which is roughly the
mass of Saturn. We emphasize that the inner and middle planet classes are likely
to undergo substantial collisional evolution during and after the protoplanetary disc phase
\citep{Lambrechts+etal2019} and that cold gas giants may be scattered into hot and warm orbits; both these effects are ignored here.

The top panel of Figure \ref{f:map} shows the results of the pebble accretion
calculations with a large initial disc size of $R_1=112\,{\rm AU}$. Here, the
formation of cold gas giants is possible up to a turbulence range between $\delta = 3 \times 10^{-5}$ and $\delta =10^{-4}$
-- and that requires Stokes numbers in the optimum range between ${\rm St}=0.01$ and ${\rm St}=0.03$.
Pebbles of higher Stokes number drift towards the star too rapidly, so that there
is insufficient time to grow cold gas giant planets before the disc has been drained of its
pebbles. As ${\rm St}$ is decreased, so is the critical turbulence strength,
with the threshold value to form cold gas giants reaching an ultra low value of $\delta=10^{-6}$ for ${\rm St}=0.001$. The formation of
super-Earths and sub-Neptunes has less strict requirements on the turbulence
strengths than the formation of cold gas giants, with a relatively high turbulence strength of
(at least) $\delta=3\times 10^{-4}$ allowed for optimum spot range in Stokes number (between ${\rm
St}=0.01$ and ${\rm St}=0.03$). Temperate rocky planets follow
super-Earths quite closely in how the limiting turbulence strength depends on
the Stokes number, although three times higher turbulence levels are allowed for forming rocky planet embryos at ${\rm St}=0.003$ and ${\rm St}=0.001$ relative to the threshold for super-Earth formation.

The lower panel of Figure \ref{f:map} maps the results when using a more nominal
initial protoplanetary disc size of $R_1=55\,{\rm AU}$. The formation of cold
gas giants now requires that the turbulence strength is at most $\delta=10^{-5}$ at an optimal value of ${\rm
St}=0.03$. The threshold for super-Earth formation remains within the strength of turbulence measured for the vertical shear instability ($\delta = 3 \times 10^{-4}$) for ${\rm St}=0.01$ to ${\rm St}=0.03$. These reasonable formation parameters for super-Earths within a protoplanetary with a nominal initial size are in agreement  with the approximately 50\% occurrence rate of planets with radii between one and four times the radius of Earth in hot and warm orbits around solar-type stars \citep{Marcy+etal2014}. The threshold for super-Earth formation falls by a factor of three compared to the larger disc for ${\rm St} \leq 0.003$. The formation of substantial rocky planet embryos requires $\delta \leq 10^{-4}$ in the smaller disc, due to the faster draining of the pebble flux and hence enhanced migration compared to the larger disc.

\section{Summary}
\label{s:summary}

In this paper we have explored how planetesimal formation by the
streaming instability and planetary growth by pebble accretion depend on the turbulence strength, pebble Stokes number and initial disc size. We summarize here a number of salient points raised by our study:
\begin{enumerate}
  \item {\it Pebble accretion can operate directly on the most massive planetesimals formed by the streaming instability.} We demonstrated analytically that the characteristic mass of planetesimals formed by the streaming instability is approximately 1/10th of the threshold mass above which pebble accretion becomes efficient, if planetesimals form at the early stages of protoplanetary disc evolution with a Stokes number of order ${\rm St} \sim 10^{-2}$. We also showed that the characteristic planetesimal mass and the pebble accretion threshold mass scale similarly with many properties of the protoplanetary disc, so that these dependencies cancel out when considering their ratio, with the exception of the Stokes number (which to first order does not seem to affect the characteristic mass of the planetesimals) and the self-gravity parameter $\varGamma$ (which does not affect the onset of pebble accretion). The resemblance of the characteristic mass of planetesimals formed by the streaming instability and the pebble accretion threshold mass implies that pebble accretion can operate efficiently already on planetesimals that are born in the upper, exponential end of the initial mass function \citep{Schafer+etal2017,Schafer+etal2024}. The initial mass function of planetesimals formed by the streaming
  instability is nevertheless not completely understood. Particularly, the mass of the largest
  bodies could be underestimated due to limited simulation
  domains \citep{Schafer+etal2024} or overestimated due to artificial merging of clumps at low resolution \citep{Johansen+etal2015,Simon+etal2016}. We also ignore an early, rapid growth phase by
  pebble accretion within the dense pebble mid-plane layer that may boost the
  masses of the largest planetesimals to values well above their initial birth
  masses. We also emphasize that while giant impacts are likely too rare in the outer disc to assist pebble accretion in forming the cores cold gas giants \citep{LorekLambrechts2026}, accretion of planetesimals and protoplanets is expected to be a major contributor to planetary growth in the inner disc, even though pebble accretion operates well there already on the planetesimal birth sizes. \\ \\

  \item {\it Migration matters for cold gas giants.} The formation of a cold gas giant akin to Jupiter requires both weak turbulence ($\delta \lesssim 10^{-4}$) and a very large disc size ($R_1 \gtrsim 100\,{\rm AU}$). These tough requirements on the
  formation of cold gas giants found here may shed light on why this class of planets is relatively rare at solar metallicity \citep{Fulton+etal2021}: their formation is only possible within the $\sim$10\% largest and most massive protoplanetary discs. We nevertheless demonstrated that reducing the amount of gas in the disc while maintaining the amount of dust leads to colder final orbits of the outermost gas giant planets. Removal of moderate amounts of gas by discs winds could thus be key to preventing gas giants from migrating to the terrestrial planet zone \citep{ZhaoMatsumura2025}. The speed of Type I migration could also be reduced by considering more sources of positive torques from the gas onto the growing protoplanet, not included here, such as the dynamical corotation torque \citep{Ndugu+etal2021,SavvidouBitsch2023}. We have also ignored pressure bumps in the disc in order to study the  limits to planet
  formation unaided by disc substructures. Pressure bumps nevertheless do appear naturally in
  simulations of MHD turbulence in the outer disc regulated by ambipolar
  diffusion \citep{Bethune+etal2016,XuBai2022,Eriksson+etal2026} and in the inner disc regulated by the Hall effect \citep{Krapp+etal2018}, by viscosity transitions in the disc, \citep{Lyra+etal2008b,ChatterjeeTan2014}, by streamer infall
  \citep{Zhao+etal2025} and by planet
  torques \citep{Lyra+etal2009,Lau+etal2024}. The increased pebble density in a pressure bump can significantly
  boost the growth rate of distant cores and thus park the final planet more easily into cold or even ultra-cold orbits \citep{JiangOrmel2023}. 
  \\ \\
  \item {\it Ice giants are hard to form even with pebble accretion.} Ice giants  akin to Uranus and Neptune in very cold orbits are even harder to form than cold gas giants under all conditions explored
  here \citep[as also concluded by][]{Eriksson+etal2023}, but the Solar System's two ice giants
  likely migrated outwards by scattering planetesimals after the dissipation of
  the protoplanetary disc \citep{FernandezIp1984,Malhotra1993} and could
  therefore have formed further in where pebble accretion rates are higher.
  Forming ice giants in the 15--20 AU region may nevertheless still require a
  gas-depleted disc to reduce their migration path. \\ \\
  \item {\it Super-Earths and sub-Neptunes form under a wide range of conditions.} These types of planets occur around approximately 50\% of solar-type stars \citep{Marcy+etal2014} and planet formation theory should therefore be able to form them straightforwardly under nominal protoplanetary disc conditions. We show that super-Earths and sub-Neptunes form in both large and mid-sized protoplanetary discs and that they tolerate a turbulence level similar to what is measured for the vertical shear instability \citep[$\delta \sim 3 \times 10^{-4}$,][]{Schafer+etal2025,HuangBai2025}. More simulations will be required in the future -- under turbulence driven by both the vertical shear instability with realistic gas cooling times set by the dust size distribution \citep{Pfeil+etal2023} and by the magnetorotational instability under the influence of Hall diffusion, Ohmic
  diffusion and ambipolar diffusion -- in order to understand how  various realistic sources of gas turbulence interacting with the streaming instability affect pebble dynamics and thus pebble accretion rates of super-Earths and sub-Neptunes growing in the inner disc.
  \\ \\
  \item {\it Wet rocky planet embryos emerge readily around the water ice line.} We distinguish in this study the parameter for angular momentum transport ($\alpha$) from the parameter characterizing the turbulence level in the disc ($\delta$), motivated by disc wind models that demonstrate cold accretion through the inner disc driven by discs winds \citep{Mori+etal2019,Mori+etal2021}. The water ice line is therefore well interior of 1 AU for most the life-time of the protoplanetary disc, when the turbulence level is weak. Planets that grow just exterior of the water ice line nevertheless sublimate the icy mantles from their accreted pebbles once these planets reach masses above 1\% of an Earth mass \citep{Johansen+etal2021}. Our rocky planets embryos growing around 1 AU are therefore endowed with water mass fractions at the level of $\sim$$0.1\%$--$1\%$, which is similar to estimates of the Earth's water budget including a majority of water residing in the core \citep{Li+etal2020}. Our water estimates nevertheless do not take into account destruction of water by reaction with Fe or Si \citep{Johansen+etal2023a,Werlen+etal2025}. We also ignored protoplanet and planet collisions that are highly probable to occur in
  the inner 10 AU both during and after the gas disc phase \citep{Lambrechts+etal2019}; we therefore clearly underestimate planetary growth as well as water budgets for the planets that form in the inner disc. Future $N$-body simulations can be used straightforwardly to include collisions while tracking the water composition, but at
  the cost of much higher computation times and more stochasticity in the results that
  will require running multiple random initial conditions for each disc model \citep{Bitsch+etal2019,Izidoro+etal2021,Yzer+etal2025}. \\ \\

\end{enumerate}
Future progress on understanding the formation of planetary systems will also benefit from better observational constraints on initial protoplanetary disc masses and radii \citep[e.g.][]{Ohashi+etal2023} as well as their temporal evolution \citep[e.g.][]{Tychoniec+etal2020,Zamudio-Ruvalcaba+etal2025}. Also, future facilities such as SKA \citep{Ilee+etal2020}, ngVLA \citep{Ricci+etal2018} and an extended ALMA \citep{Burrill+etal2022} will be key to probing conditions in the inner regions of protoplanetary discs, such as detecting substructures caused by forming super-Earths there, measuring the sizes and Stokes numbers of inner-disc pebbles as well as putting constraints on turbulence levels in the disc regions where the vertical shear instability is expected to be active in concert with the streaming instability.

\begin{acknowledgements}
We thank the anonymous referee for insightful comments that helped improve the manuscript. A.J.\ acknowledges funding from the Carlsberg Foundation (Semper Ardens: Advance grant FIRSTATMO). W.L.\ acknowledges funding from NASA Emerging Worlds program via grant No. 80NSSC22K1419 and NSF via grant 2511672. W.L. further acknowledges the hospitality of the Center for Star and Planet Formation at the University of Copenhagen during a sabbatical leave, which facilitated discussions that contributed to this paper.
\end{acknowledgements}

\end{document}